\documentclass[12pt]{article}

\usepackage{natbib}
\usepackage{epsfig}

\begin{document}

\renewcommand{\refname}{\normalsize \bf \em References}

\title{\bf LAYER-BY-LAYER GROWTH FOR PULSED LASER DEPOSITION}
\author{
B.\ HINNEMANN\footnote{
      Corresponding author. Tel.\ +49-203-3793321;
      fax: +49-203-3791681. 
      \newline E-mail: hinne@comphys.uni-duisburg.de},
F.\ WESTERHOFF and D.E.\ WOLF
\\*[0.2cm]
    {\small \it Theoretical Physics,} \\
    {\small \it Gerhard Mercator University, 47048 Duisburg, Germany}
\\*[0.2cm]
}
%
\maketitle 
%
\begin{abstract}
%
Pulsed laser deposition (PLD) is a popular growth method, which has been successfully used for fabricating thin films. Compared to continuous deposition (like molecular beam epitaxy) the pulse intensity can be used as an additional parameter for tuning the growth behavior, so that under certain circumstances PLD improves layer-by-layer growth. We present kinetic Monte-Carlo simulations for PLD in the submonolayer regime and give a description of the island distance versus intensity. Furthermore we discuss a theory for second layer nucleation and the impact of Ehrlich-Schwoebel barriers on the growth behavior. We find an exact analytical expression for the probability of second layer nucleation during one pulse for high Ehrlich-Schwoebel barriers. 
%
\end{abstract}


\section*{1. INTRODUCTION}
Pulsed Laser Deposition (PLD) is a growth method increasingly used for the fabrication of thin films [\cite{chrisey:book}]. It is especially suited for the growth of complex multicomponent thin films, e.g. high temperature superconductors [\cite{cheung:paper}], biomaterials [\cite{cotell:paper}], or ferroelectric films [\cite{ramesh:paper}]. A great advantage of PLD is the conservation of the stoichiometry of virtually any target material in the deposition.

The main feature of PLD is that the target material is ablated by a {\em pulsed} laser and then deposited onto the substrate. Thus, in one pulse many particles arrive at the surface simultaneously. The time between two pulses is of the order of seconds and the pulse length usually is of the order of nanoseconds [\cite{cheung:review}]. Therefore we are going to neglect the pulse length.

In this paper we present computer simulations showing several features of the growth morphology in PLD. In the next section a short description of the typical island distance for PLD in the submonolayer regime will be given. Thereafter the growth behavior of PLD under the influence of Ehrlich-Schwoebel barriers will be discussed. The aim of this work is to give an insight into possible reasons why in some experimental situations PLD has been shown to produce better layer-by-layer growth than ordinary molecular beam epitaxy (MBE) [\cite{jenniches:paper}].

\section*{2. ISLAND DISTANCE}
\begin{figure}
\begin{center}
\epsfig{file=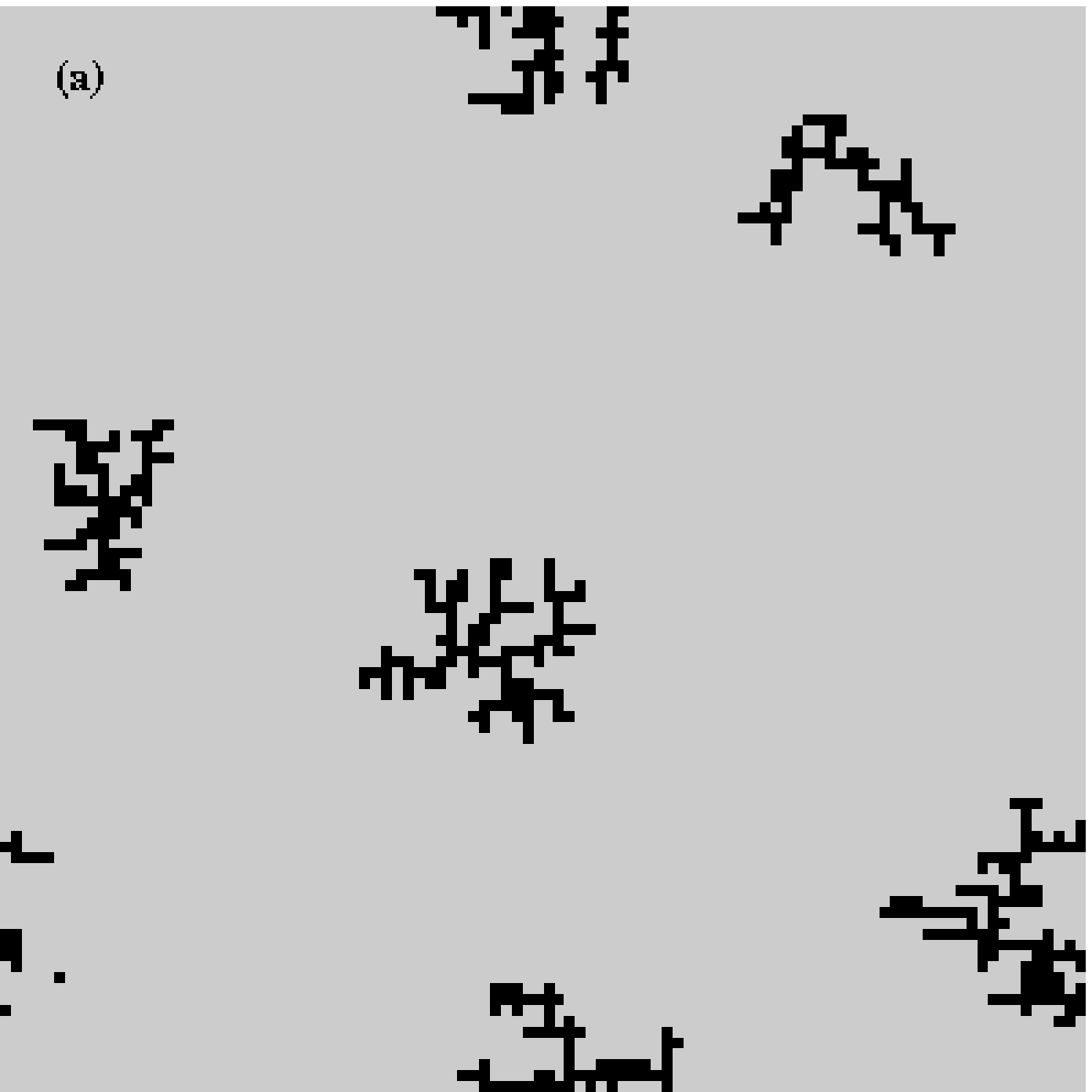,width=5cm, angle=0}
\hspace{0.2cm}
\epsfig{file=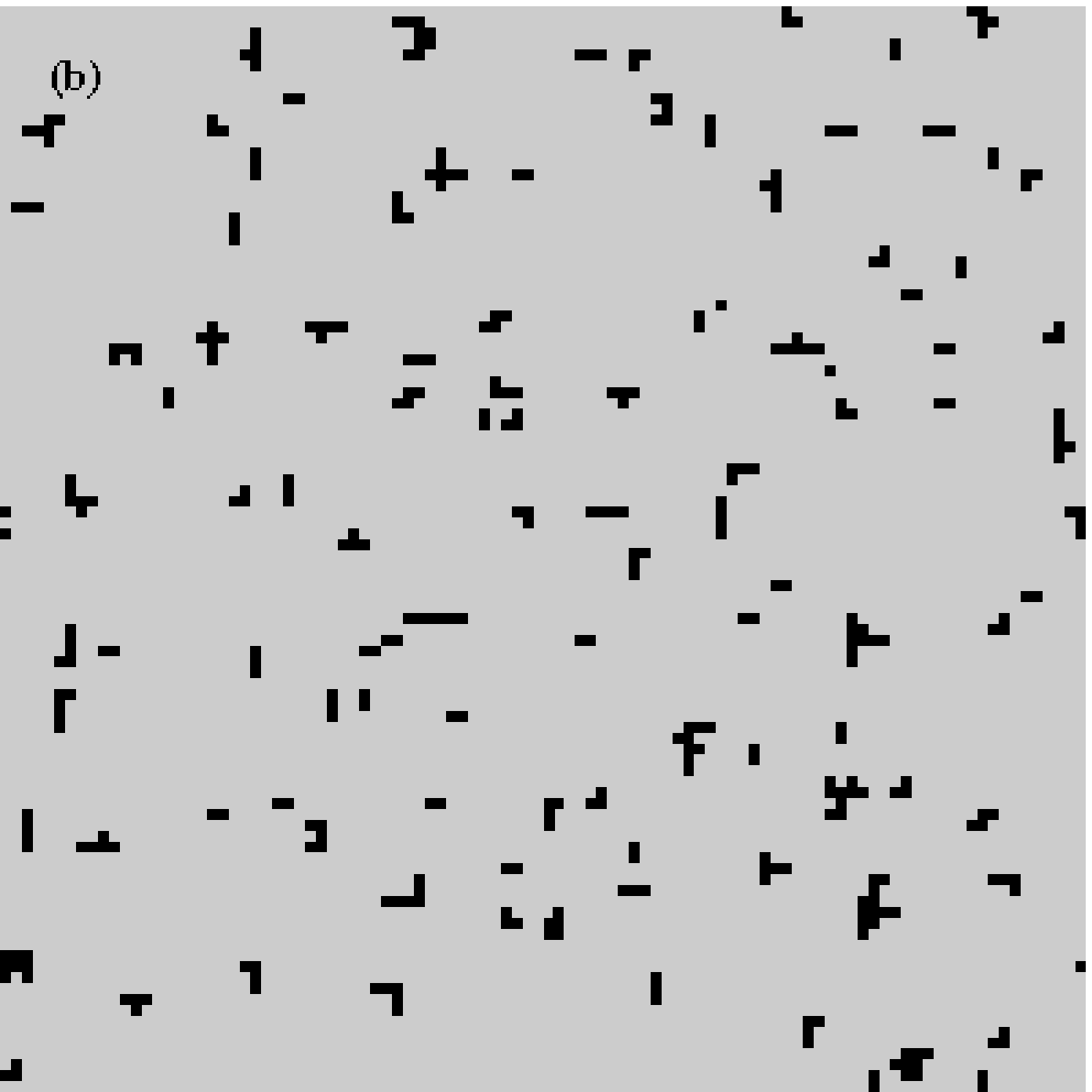,width=5cm, angle=0}
\caption{(a) A simulated MBE-grown surface with $D/F=10^8$. (b) A simulated PLD-grown surface with $D/F=10^8$ and $I=0.01ML$. The coverage of both surfaces is $0.05ML$. }
\label{fig:MBEcompare}
\end{center}
\end{figure}

\begin{figure}
\begin{center}
\epsfig{file=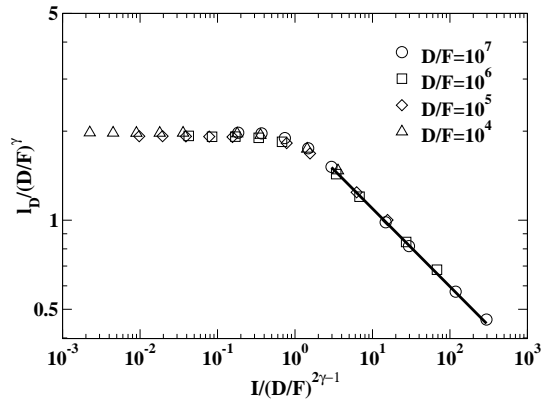,width=8cm, angle=0}
\caption{The scaled island distance for PLD versus the scaled intensity.}
\label{fig:scalcoll}
\end{center}
\end{figure}

The control parameters of PLD are the intensity $I$, which is the number of particles deposited in one pulse per unit area, and the diffusion-to-deposition ratio $D/F$. The average deposition rate is given by $F=I/\Delta t$, where $\Delta t$ is the time interval between two pulses. The intensity is measured in monolayers (ML), and $D/F$ is dimensionless, as the lattice constant is set to unity. The surface morphology depends sensitively on the pulse intensity which qualitatively can be seen in fig. \ref{fig:MBEcompare}. One notices that in MBE there are few large islands relatively far apart whereas for PLD with a high intensity the surface is covered with many small islands. 
However, if one reduces the intensity to one particle per pulse, one would expect PLD to produce the same island morphology as MBE. This is indeed the case, although the two situations are not exactly the same, as in PLD the deposition takes place at fixed times whereas in MBE it is probabilistic and therefore the time intervalls between two depositions have a Poissionan distribution. This difference, however, only influences the island morphology at high $D/F$, where finite-size effects are setting in [\cite{hinnemann:thesis}]. In the simulations presented here only $D/F$ below the finite-size region are considered. The simulations have been performed on a $400\times 400$ square lattice and the island distance has been measured at $0.2ML$ coverage, when the island density reaches its maximum but coalescence does not yet set in.
In the following the dependence of the island distance on the ratio $D/F$ and on the pulse intensity $I$ is investigated. For small intensities we recover the well-known power law for the island distance in MBE
\begin{equation}
l_D\propto \left(\frac{D}{F}\right)^{\gamma}
\label{eq:mberegime}
\end{equation}
with the exponent $\gamma$ depending on the dimension of the surface, the island dimension, and the critical nucleus $i^*$, i.e. the smallest stable island contains $i^*+1$ atoms [\cite{stoyanov:paper}; \cite{schroederwolf}]. For a two-dimensional surface, compact islands and a critical nucleus of \mbox{$i^*=1$}, one obtains $\gamma =1/6$. One should note that here the islands are not compact but fractal, as can be seen in fig. \ref{fig:MBEcompare}. This is due to the fact that edge diffusion is not considered in the simulations. The exponent $\gamma$ can be determined from simulations such, that one 
monitors the number of nucleation events in a layer ($\propto l_D^{-2}$) as a function of $D/F$.
The value obtained in the present simulations is 
$\gamma=0.17\pm0.01$. For PLD with large intensities the island distance 
obeys a different power law, however:
\begin{equation}
l_D\propto I^{-\nu}.
\label{eq:pldregime}
\end{equation}
In this regime the island distance is independent of the parameter $D/F$, since the adatoms do not make use of their diffusion probability, as they find an island and attach to it in a much shorter time as the time they are allowed to diffuse between two depositions. The two regimes described by (\ref{eq:mberegime}) and (\ref{eq:pldregime}) are separated by a crossover at a certain intensity, where the number of deposited atoms is of the same order of magnitude as the adatom density. As the average adatom density $n$ in MBE scales as $n\propto\left(D/F\right)^{-1+2\gamma}$, the critical intensity has to show the same scaling behavior
\begin{equation}
I_c\propto \left(\frac{D}{F}\right)^{-1+2\gamma}.
\label{eq:critintensity}
\end{equation}
It follows from (\ref{eq:mberegime}), (\ref{eq:pldregime}) and (\ref{eq:critintensity}) that the island distance can be represented as a scaling law [\cite{westerhoff:paper}]
\begin{equation}
l_D\propto \left(\frac{D}{F}\right)^{\gamma}\cdot f\left(\frac{I}{I_c}\right)
\label{eq:scaling}
\end{equation}
where $f$ is a scaling function with the asymptotic behavior
\begin{equation}
f(y)\left\{\begin{array}{ll}
                =\mbox{const.} & \mbox{for } y\ll 1\\
                \sim y^{-\nu}  & \mbox{for } y\gg 1.
           \end{array}\right.
\end{equation}
Since $l_D$ does not depend on $D/F$ for high intensities, the factor $(D/F)^{\gamma}$ in (\ref{eq:scaling}) must be compensated by the $I_c$-dependence of $f$:
${I_c}^{\nu}\cdot (D/F)^{\gamma}=\,\mbox{const.}$, which together with (\ref{eq:critintensity}) leads to
\begin{equation}
\nu =\frac{\gamma}{1-2\gamma}.
\label{eq:nu}
\end{equation}
The data collapse of the simulation results according to (\ref{eq:scaling}) is shown in fig. \ref{fig:scalcoll}. The exponent $\nu$ 
obtained from the slope in the double-logarithmic plot is
$\nu=0.26\pm0.01$ and thus in agreement with the predicted value $\nu =1/4$. Furthermore fig. \ref{fig:scalcoll} shows that the island distance indeed obeys the proposed scaling law. Thus there is a critical intensity which divides the intensity parameter space in MBE-like and different behavior. Below the critical intensity, the island distance in PLD equals or is very similar to the one in MBE, whereas above the critical intensity it differs significantly.

\section*{3. GROWTH WITH EHRLICH-SCHWOEBEL BARRIERS}
\begin{figure}
\begin{center}
\epsfig{file=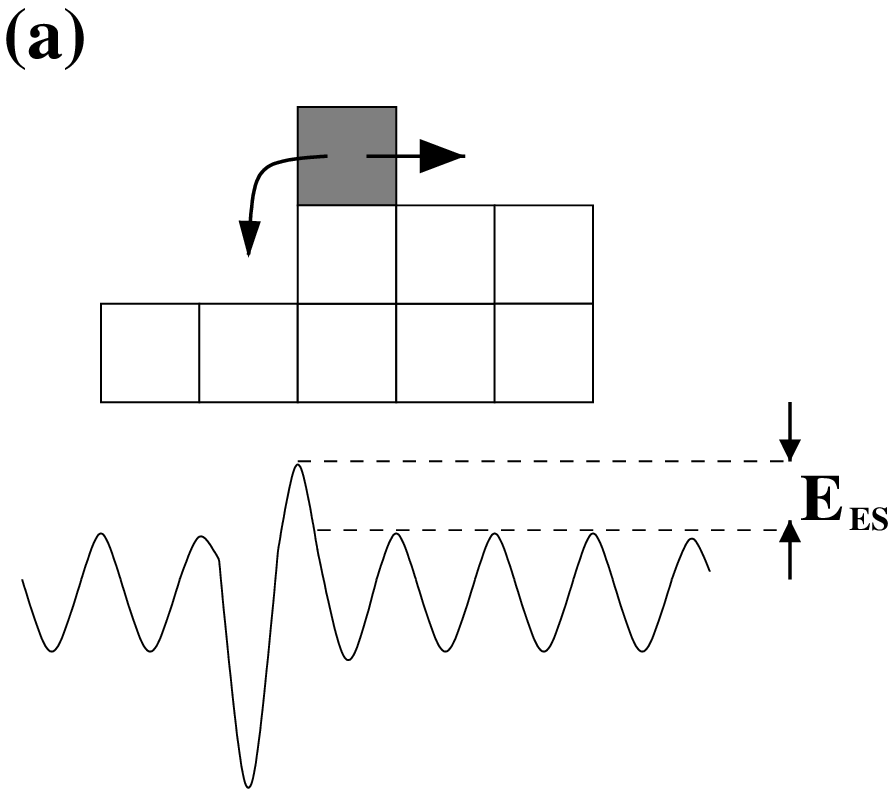,width=5cm, angle=0}
\hspace{0.2cm}
\epsfig{file=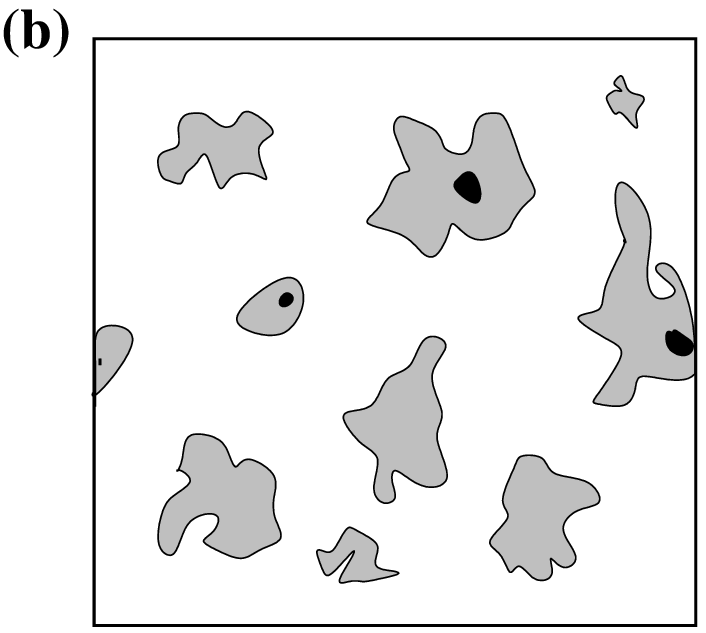,width=5cm, angle=0}
\caption{(a) Illustration of the Ehrlich-Schwoebel effect. The atom at the edge(black) has to overcome an additional energy barrier $E_{ES}$ to hop down. (b) A model surface at the onset of second-layer nucleation. The calculations are done in the limit $D/F\rightarrow\infty$.}
\label{fig:schwoebel}
\end{center}
\end{figure}

In most experimental situations of PLD and MBE there are barriers to interlayer transport present, i.e. an atom experiences an extra barrier in addition to the diffusion barrier when hopping down from an island [\cite{ehrlichhudda,schwoebel:paper}]. This barrier is termed Ehrlich-Schwoebel barrier (ES-barrier) and is of the order of $\sim 0.1eV$ for metals [\cite{ruggerone:paper}]. Intuitively the ES-barrier can be explained as follows: In order to jump down an island edge, the atom 
goes through a position with a lower coordination than at the saddle point of diffusion on an island. The ES-barrier $E_{ES}$ is illustrated in fig. \ref{fig:schwoebel}. A very useful measure for its effect is the Schwoebel length, which is defined as
\begin{equation}
l_{ES}\equiv exp\left(\frac{E_{ES}}{k_BT}\right)
\label{eq:schwoebel}
\end{equation}
and will be used as the control parameter in the following. The Schwoebel length is dimensionless, as the lattice constant is set to unity. In general the Ehrlich-Schwoebel barrier impedes layer-by-layer growth, as it forces the atoms to stay on the islands for a longer time. This increases the probability that two atoms on an island meet and nucleate before they can leave the island separately. A model surface with second layer nucleation in the submonolayer regime is shown in the right panel of fig. \ref{fig:schwoebel}.
In the following a description of the growth with ES-barriers in the submonolayer regime is given. It is important to describe second layer nucleation as it contributes to surface roughening and therefore has to be reduced as much as possible in order to achieve layer-by-layer growth. This concept of formulating a second layer nucleation theory and predicting the growth mode for a given parameter set has been successfully applied to MBE [\cite{tersoff:paper}], and can be generalized to PLD as follows. 

Whether one obtains layer-by-layer growth depends on the time during deposition of the first monolayer until nucleation in the second layer starts. If second layer nucleation does not start significantly before the first layer is completed, one gets layer-by-layer growth, but if it starts well before the completion of the first monolayer, the surface will roughen [\cite{villain1991}]. If one compares PLD to MBE under these circumstances, the high intensity, which produces small islands, seems advantageous, as adatoms on the islands can leave them more quickly, provided they are small enough. On the other hand a high intensity means, that many particles are deposited on the surface simultaneously, which increases the probability for second layer nucleation, as two particles may be deposited on the same island and meet before they can both leave the island separately.

Now a description for the {\it second layer nucleation probability} $p_{nuc}$ in PLD will be given. Let this quantity be defined as the probability that on a given island of size $A$ at least one nucleation occurs during one pulse. As it has been shown in a similar analysis for MBE, the probability of second layer nucleation can be used as a starting point of a second layer nucleation theory [\cite{krug:paper}]. It will be calculated making the following assumption: If two or more atoms are deposited simultaneously on the same island, they meet and nucleate before they leave the island. This assumption is true for high enough Ehrlich-Schwoebel barriers. Moreover we assume that in this case the nucleation happens before the next pulse arrives, which requires that $D/F$ is high enough. Then the probability of second layer nucleation equals the probability that two or more atoms are simultaneously deposited on an island of size $A$ during a pulse of intensity $I$. Thus we obtain for $D/F \rightarrow \infty$:
\begin{equation}
p_{nuc}=1-\exp{(-IA)(1+IA)},
\label{eq:nucprobability}
\end{equation}
where $A$ is the island area.
In order to verify (\ref{eq:nucprobability}), simulations have been performed, where the second layer nucleation probability was measured in dependence of the island size. The simulations were performed on a $400\times 400$ lattice for various intensities $I$ and Ehrlich-Schwoebel barriers, whose size is indicated by the Schwoebel length (\ref{eq:schwoebel}). 
\begin{figure}
\begin{center}
\epsfig{file=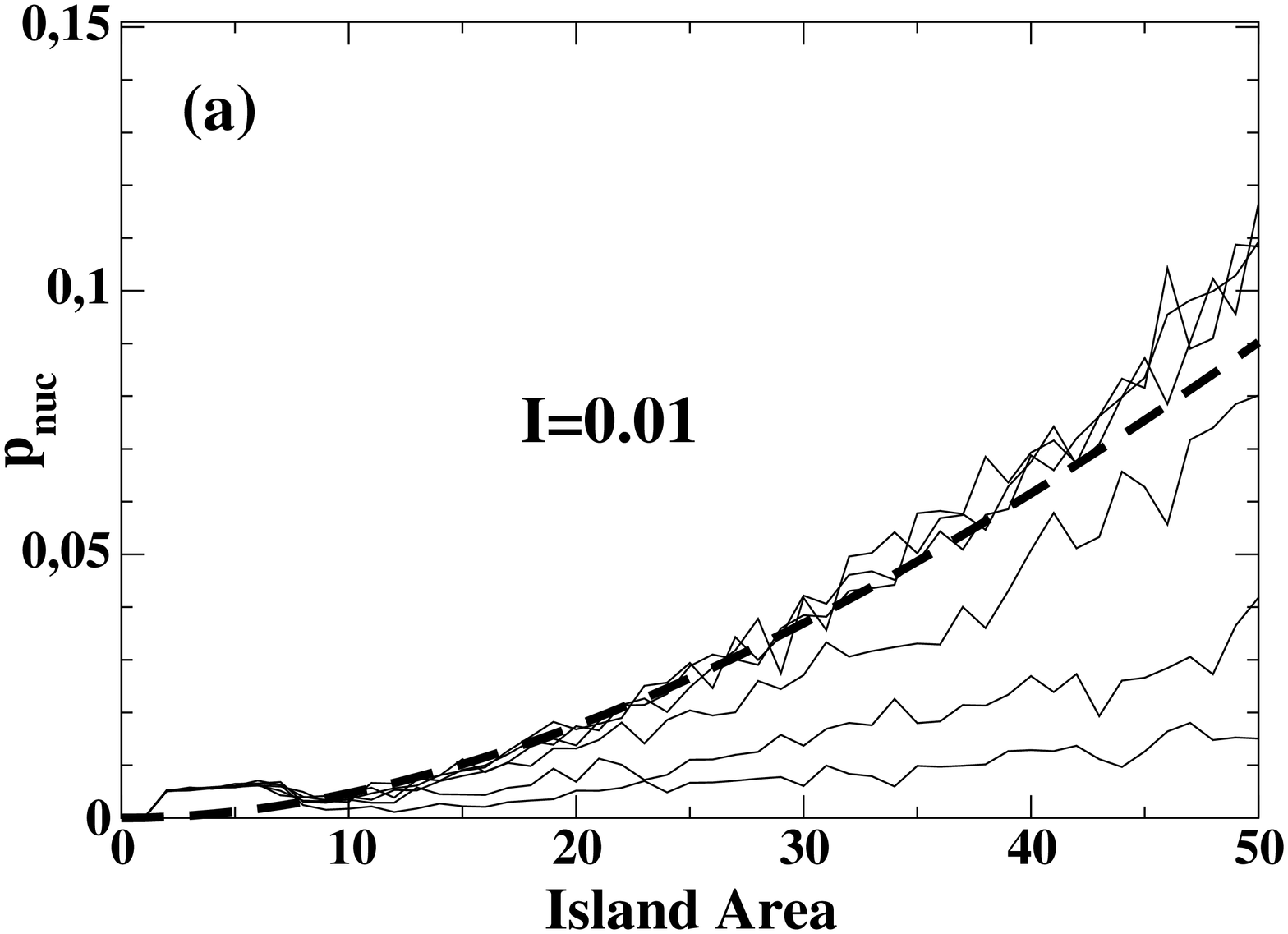,width=4.7cm, angle=270}
\hspace{0.2cm}
\epsfig{file=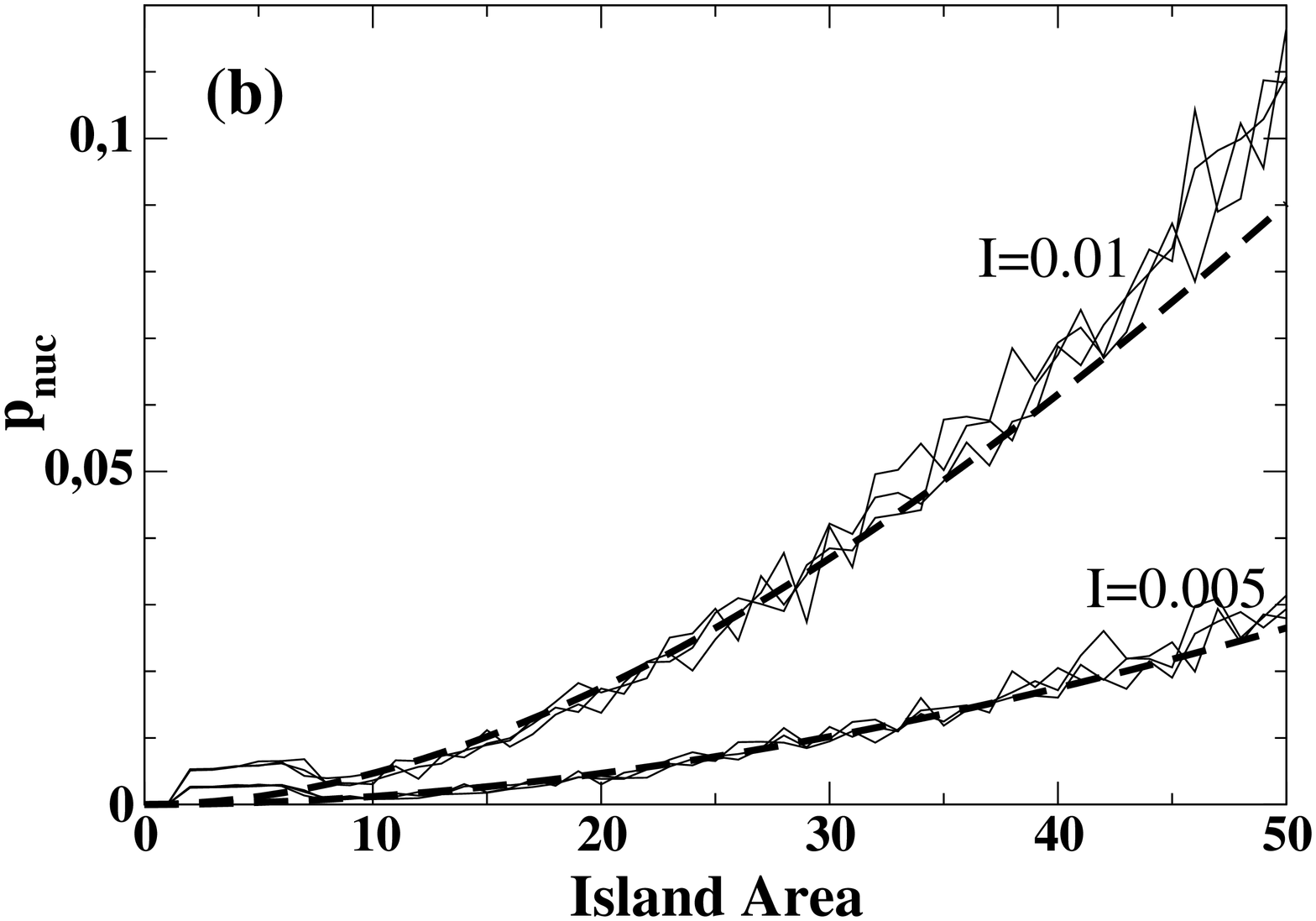,width=4.7cm, angle=270}
\caption{(a) The second layer nucleation probability $p_{nuc}$ for $I=0.01$ versus island size $A$ for different Schwoebel-barriers. From bottom to top the Schwoebel lengths are $l_{ES}=10^0,10^1,10^2,10^3,10^4,10^5$. (b)  Plot of $p_{nuc}$ versus $A$ for two different intensities and for high Schwoebel barriers $L_{ES}=10^3, 10^4, 10^5$.}
\label{fig:nucprobability}
\end{center}
\end{figure}

In fig. \ref{fig:nucprobability} to the left the results for $p_{nuc}$ are displayed for the intensity $I=0.01$ and for different Schwoebel-lengths. One can see that the measured curves approach the predicted curve for increasing Schwoebel-barriers and that the agreement is good for $l_{ES}\geq10^3$, when the Schwoebel-barrier is high enough to fulfill the assumption that atoms deposited on an island meet and nucleate instead of leaving it. The agreement is not convincing for island areas below $10$, as here the discrete nature of the island sites plays a role. The measured curves also deviate from the predicted curve for island sizes above $40$. These deviations are due to coalescence, which is not accounted for in the calculation of $p_{nuc}$ and which for $I=0.01$ already starts for island sizes below $50$. This explanation is supported by the diagram in fig. \ref{fig:nucprobability} to the right. The measured curves and the predicted curve for $I=0.005$ agree much better, as for a lower intensity coalescence starts at a larger characteristic island size. In general one can see that the second layer nucleation probability $p_{nuc}$ can be described adequately with (\ref{eq:nucprobability}) for island sizes smaller than the characteristic island size at the onset of coalescence.

\section*{4. CONCLUSIONS}
The island distance depending on the intensity $I$ and $D/F$ has been described by a scaling law. For intensities larger than $I_c$ the typical island area 
scales like $A \propto I^{-2 \nu}$.
Inserting this into (\ref{eq:nucprobability}) leads to the conclusion that pulsed laser deposition in the limit $D/F \rightarrow \infty$ should give worse results than molecular beam epitaxy for layer-by-layer growth. This is clear, as molecular beam epitaxy leads to layer-by-layer growth in this limit even for high, but finite ES-barriers. We emphasize, however, that strain effects and transient mobility of freshly deposited atoms have not been taken into account in the present investigation. 

For finite $D/F$ the situation is less clear.
In order to further analyze this aspect, one needs a description of the time-dependent island area which will be published elsewhere [\cite{hinnemann:tobepublished}]. 
If one compares the growth mode of PLD and MBE, one can already infer, that the use of PLD can only be advantageous for materials with a relatively high Ehrlich-Schwoebel barrier, as there the small islands have the advantage that atoms deposited on them can leave them with a higher probability before they nucleate on them. For growth situations with a very small Schwoebel-barrier, the diminished island size is no advantage, on the contrary, as it enhances second-layer nucleation compared to thermal deposition. The conclusion is that it does not help to just increase the intensity in order to obtain better growth results with a high Schwoebel-barrier. At some point the increased second layer nucleation probability outweights the advantage of small islands [\cite{hinnemann:thesis}].



\end{document}